\documentclass[sigconf, nonacm]{acmart}

\usepackage{tcolorbox}
\usepackage{bm}
\usepackage{pifont}
\usepackage{tikz}
\usetikzlibrary{positioning, arrows.meta, shapes.geometric, fit, backgrounds, calc}
\usepackage{listings}
\usepackage{enumitem}

\definecolor{codeblue}{rgb}{0.0, 0.3, 0.6}
\definecolor{codegray}{rgb}{0.5, 0.5, 0.5}
\definecolor{codegreen}{rgb}{0.0, 0.5, 0.2}
\definecolor{lightblue}{RGB}{220, 235, 252}
\definecolor{lightorange}{RGB}{255, 240, 220}
\definecolor{lightgreen}{RGB}{220, 245, 220}
\definecolor{lightpurple}{RGB}{235, 225, 250}

\begin{document}

\title{DataClaw: An Autonomous Data Agent with Instant Messaging Integration}

\author{Huahang Li, Wentao Hu, Zhuoyue Wan, Chen Jason Zhang, Haoyang Li, Xiaoyong Wei}
\affiliation{%
  \institution{The Hong Kong Polytechnic University, Hong Kong, China}
}
\email{{{hua-hang.li, wayne-wt.hu, zhuoy.wan}@connect.polyu.hk,{jason-c.zhang, haoyang-comp.li, cs007.wei}@polyu.edu.hk}}

\vspace{3em}
\begin{abstract}

In daily life, there are many scenarios that people need to tackle data-related tasks, such as filling out forms, analyzing Excel files, and visualize data report. However, the tools available for these tasks often fragment, requiring users to switch between multiple applications and manually orchestrate steps like data processing, querying, and visualization. Moreover, these tools often assume a certain level of technical proficiency, creating barriers for non-technical users. 
To facilitate tacking daily data task, we present DataClaw, an autonomous data agent that integrates directly into familiar instant messaging (IM) platforms. By simply typing a natural language request in a chat interface, users enable DataClaw to autonomously plan and execute a complete analytical pipeline, delivering insights, charts, and reports directly back into the conversation. Under the hood, DataClaw is powered by a transparent ReAct reasoning engine, a multi-tiered memory system for cross session context preservation, and a pluggable skill architecture for on-the-fly extensibility.
In this demonstration, attendees will interact with DataClaw via standard IM platforms to solve real-world data scenarios, experiencing how it serves as a highly capable personal data assistant.
\end{abstract}

\maketitle

\section{Introduction}

\begin{figure}[t]
  \vspace{3em}
  \centering
  \includegraphics[width=\linewidth]{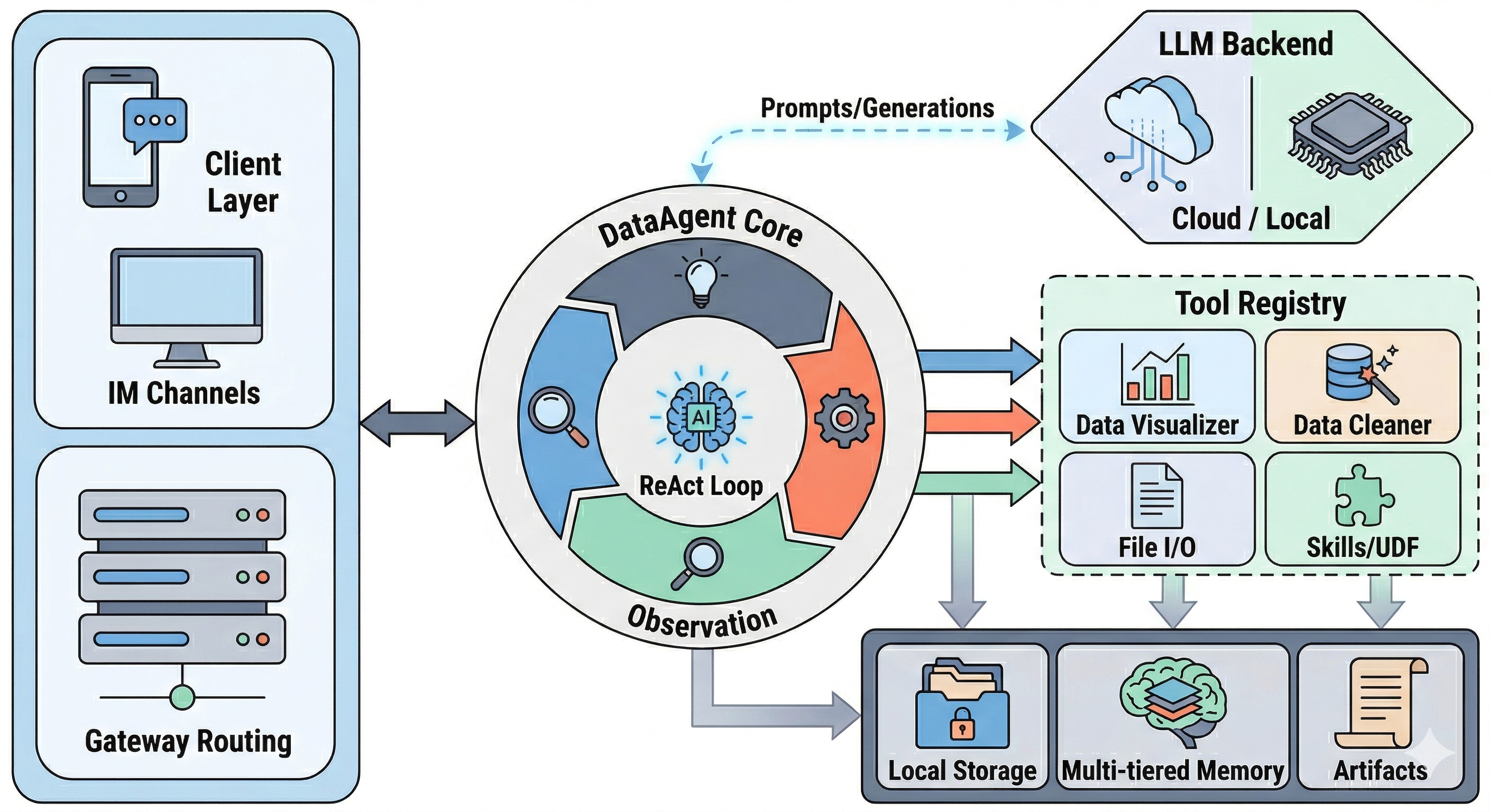} 
  \caption{The architecture of DataClaw.}
  \label{fig:architecture}
  \vspace{-1em}
\end{figure}

Living in the era of information, everyday people work with data in various forms, from analyzing sales figures and processing customer information to filling out forms. However, the tools available for these tasks often fragment the workflow, requiring users to switch between multiple applications and manually orchestrate steps like data processing, querying, and visualization. This fragmentation not only reduces efficiency but also increases the risk of errors. Moreover, these tools often assume a certain level of technical proficiency, creating barriers for non-technical users who need to interact with data but may not be comfortable with complex software interfaces or coding. 

Recent advances in large language models (LLMs)~\cite{openai2023gpt4, wang2024survey} have prompted a wave of chat-based analytical assistants that promise to replace this fragmented workflow with a conversational interface. These tools are genuinely useful for short, self-contained questions, yet they carry assumptions that limit their applicability in practice~\cite{narayan2022can, luo2018deepeye, db-gpt}. Most are optimized for stateless, cloud-hosted interactions, meaning that local files must be uploaded, execution happens on remote infrastructure, and neither intermediate artifacts nor session context persists beyond a single conversation.

Traditional business intelligence (BI) software sits at the opposite end of the spectrum: rich dashboards and visual query builders, but still fundamentally requiring users to drive every step manually, assembling the pipeline themselves rather than delegating it. Neither approach achieves what a daily data solver actually needs, an agent that can accept a high-level request, plan and execute a multi-step workflow autonomously, while keep data local and deliver the result directly.

We introduce DataClaw\footnote{The project homepage and source code are avaible at \url{https://github.com/HKPolyU-IDEAL/DataClaw}. }, an autonomous data agent that help users tackle daily data tasks through natural language interactions within their familiar IM enviroment. While existing foundational frameworks like OpenClaw~\cite{openclaw2026} provide general purpose reasoning and execution environments, they lack the specialized data-centric toolchains, and system designs required for continuous data workflows. To enable an understanding personal data agent, DataClaw incorporates three key design principles. First, the ReAct execution model ensures that the agent's reasoning is auditable and self-correcting, every intermediate thought, action, and observation is visible in the console, so users can follow the agent's logic and intervene if needed. Second, a pluggable skill mechanism allows new capabilities to be added on the fly without modifying the core agent code, making it adaptable to new domains and team-specific workflows. Third, a multi-tiered memory mechanism preserves schema context, past findings, and user preferences across sessions, so the agent accumulates understanding of a dataset over time rather than starting cold on every request. Considering data governance and privacy concerns, DataClaw canexecutes entirely on a local machine or private server. It leverages lightweight execution engines to process tabular data efficiently and also supports a variety of LLM backends, including locally hosted models (i.e., Llama~\cite{meta2024llama3}). So practitioners can choose the level of autonomy and data governance that suits their needs, without uploading anything to the cloud.

\begin{figure*}[t]
\centering
\includegraphics[width=\linewidth]{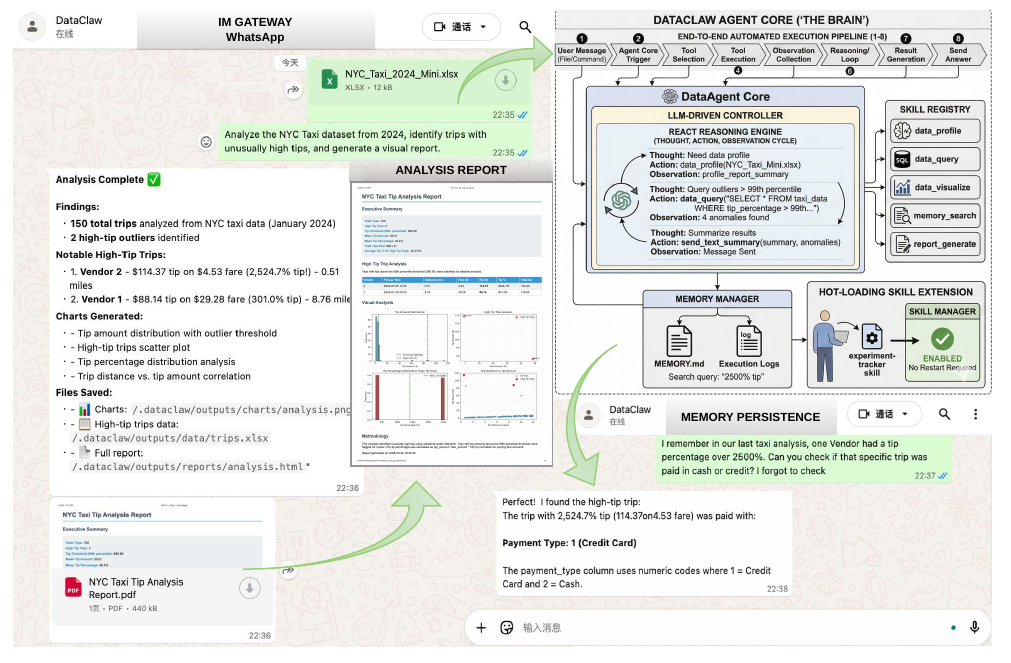}
\vspace{-2em}
\caption{DataClaw operates natively within IM platforms. Users can send natural language requests and receive results directly in the chat thread, while the backend console provides a real-time view of the agent's reasoning process.}
\label{fig:im}
\vspace{-1em}
\end{figure*}

In this demonstration, we present DataClaw as a running system and focus on the design choices that make autonomous local data assistant practical at the interface layer. We walk attendees through a sequence of live scenarios that highlight the core experience, turning a single natural-language request into a complete autonomous workflow, as well as the system's memory, continuity, and extensibility features. By the end of the demo, attendees will have seen how DataClaw transforms the way people interact with data, making personal data assistant accessible to everyone in their daily use.

\vspace{-2em}

\section{System Overview}

\noindent\textbf{Overview.} As shown in Figure~\ref{fig:architecture}, DataClaw adopts a three-tier architecture that separates user interaction, data agent core, and memory \& execution environment. At the outermost tier, users interact through familiar IM platforms or a lightweight browser console. As illustrated in Figure~\ref{fig:im}, users can simply issue requests. These messages are received by a gateway, and then passed to the data agent core, which carries out the actual work, executes a ReAct loop that draws on the skill extensions, and persistent memory modules to complete a task through a sequence of intermediate steps. Finally, the results are delivered back to the user through the same IM channel, and the conversation logs and artifacts are safely stored on the local machine for future reference.

\noindent\textbf{Supported tasks.} DataClaw is designed to handle a wide range of data-related tasks, including but not limited to:
\begin{itemize}[leftmargin=*]
  \item \textit{Data exploration}: To provide a brief understanding of the structure and content of tabular data, including schemas, data types, and basic statistics.
  \item \textit{Data profiling}: Analyze data quality and characteristics, such as completeness, consistency, and distribution.
  \item \textit{Querying}: Retrieve specific information from data sources via filtering, aggregation, and joining.
  \item \textit{Cleaning}: Identify and correct errors or inconsistencies, such as missing values, duplicates, and outliers.
  \item \textit{Form filling}: Automate data-driven form completions to ensure accuracy and reduce manual effort.
  \item \textit{Visualization \& reporting}: Create visual report and comprehensive data summaries.
\end{itemize}
 
Through the built-in tool registry, the agent has access to a curated set of operations that cover the canonical stages of a data workflow, and through the pluggable skill extension, users can extend these capabilities to new domains and specific workflows on the fly.

\section{Core Components}
\label{sec:core_components}

To enable the autonomous data assistant in Section 2, DataClaw is designed with modularity and extensibility in mind, with four core components, namely ReAct-based Reasoning Engine, Multi-Tiered Memory System, Pluggable Skills and Tool Registry, and Asynchronous Event Streaming Channel. Each component is built to be flexible and adaptable, allowing DataClaw to meet a wide range of user needs while maintaining a seamless experience. We then walk through each component in detail:

\subsection{ReAct-based Reasoning Engine}

The core of DataClaw is a ReAct-based reasoning engine that serves as the brain of the system. It analyzes user requests, plans a sequence of actions, and executes them iteratively while maintaining an auditable execution trace in the backend console. The ReAct paradigm~\cite{yao2023react} interleaves a thought, an sction, and an obbservation at every step. In our implementation, we set a maximum iteration limit (e.g., 50) to prevent infinite loops. The maximum context window is also configurable, and the agent can invoke a built-in \texttt{memory\_search} tool to retrieve relevant information from the persistent memory when needed. After each action, the agent allows to reflect on the observation and adjust its plan accordingly. This is particularly important in data tasks, where the agent may encounter issues like missing values, schema mismatches, or ambiguous queries that require iterative refinement. 

\subsection{Multi-Tiered Memory System}

Though the ReAct loop improves the agent's ability to handle complex tasks, it does not address the issue of context loss. Even for a human data analyst, it is impossible produce a perfect answer without context and prior understanding of the task. To ensure DataClaw can maintain continuity over time, we design a multi-tiered memory system that preserves both short-term conversational context and long-term knowledge about the dataset and user preferences. Technically, it is implemented as a combination of in-memory data structures and persistent markdown files. At the session level, a rolling \texttt{DataMemory} allows the agent to keep track of the current context and user interactions. However, in long-running data exploration, table schemas, error messages, and long Python outputs can easily overflow the token limit. To address this, a \texttt{MemoryCompactionHook} monitors token usage and automatically compresses older messages into concise semantic summaries once a configurable threshold is reached. This ensures that the agent can continue to operate effectively without losing critical context. For cross-session and global continuity, DataClaw maintains persistent files: \texttt{MEMORY.md} and \texttt{AGENTS.md}, which store the agent's accumulated understanding of sessions, and the gobal configuration respectively. While a central \texttt{SOULS.md} summarizes the user's preferences and personal notes over time. This mechanism avoid the cold-start each new session, allowing the agent to build a deeper understanding of the user's data and needs over time.

\subsection{ Tool Registry \& Pluggable Skills}


After the reasoning engine determines the next action, the agent needs to act on it to make progress. The Tools serve as the hands of DataClaw, which execute actions on local machine. And the SKills serve as the pre-defined procedures that can be invoked by the agent to perform specific tasks. We provide a set of bulit-in tools that cover common data operations, such as pdf parsing, Excel editing, and so on. And also Skills like data profiling, report generation, and experiment tracking. 

One key design of DataClaw is the pluggable skill mechanism, which allows users to extend these capabilities on the fly. In practice, a skill is a secure bundle of a plain-text instruction file, \texttt{SKILL.md} containing trigger keywords and few-shot examples. There could include optional Python modules to take a procedure of actions. The hot-loading mechanism enables the user to customize DataClaw to their specific needs. For example, a user could create a skill for analyzing sales data that includes specific visualization templates relevant to their business. By simply dropping the skill into the designated directory, DataClaw can instantly register the new capabilities without requiring a system restart.  

\subsection{Asynchronous Event Streaming Channel}

To provide a user-friendly experience, DataClaw incorporates an asynchronous event streaming channel that allows the agent to communicate its intermediate states back to the user in real time. This is particularly important for data tasks, which can be long-running and may involve multiple steps. By streaming updates such as \textit{thinking}, \textit{tool\_call}, and \textit{tool\_result} through Server-Sent Events (SSE), users can follow the agent's reasoning process without experiencing wait anxiety. This design also enhances transparency and trust, as users can see exactly what the agent is doing at each step and intervene if necessary.

Futhermore, the asynchronous channel is configurable to support different levels of verbosity. User can set preferences for how much information they want to receive during the execution. For instance, some users may prefer to only see final results, while others may want detailed logs of every thought and action. This flexibility ensures that DataClaw can cater to a wide range of user preferences and use cases from quick queries to in-depth analyses.

\section{Demonstration}
\label{sec:demo}
\begin{figure*}[t]
\centering
\includegraphics[width=\linewidth]{figures/UI.pdf}
\vspace{-3em}
\caption{DataClaw's console UI components include (a) the Sidebar Control Panel, (b) the Channel Config, (c) the Skills Management, (d) System Status and Logs, and (e) the Memory Management.}
\label{fig:ui}
\vspace{-1em}
\end{figure*}

In this demonstration, attendees are invited to experience a zero-install, interactive demonstration using their personal smartphones. By simply scanning a QR code at our booth via Telegram or WhatsApp, attendees can instantly connect to a virtual DataClaw agent. The demonstration walks attendees through a sequence of live scenarios on the NYC Taxi Trip dataset. We will showcase how DataClaw can handle complex analytical tasks. The backend console, as shown in Figure~\ref{fig:ui}, provides a real-time view of the process, allowing attendees to see the agent's reasoning and tool usage.

\textbf{1. End-to-End Analysis via IM.} 
The core experience begins with an attendee sending a request, such as \textit{``Analyze the NYC Taxi dataset for high tips and generate a visual report.''} DataClaw autonomously decomposes this goal, profiling the data (\texttt{data\_describe}), issuing targeted queries (\texttt{data\_query}), and producing charts without demanding SQL or scripting. In real time, the ReAct loop renders its thought, action, and observation traces on the backend console for full auditability. Seconds later, the user receives charts picture and a PDF report directly in the chat thread, with all artifacts safely persisted on the local machine.

\textbf{2. Memory, Continuity, and Extensibility.} 
To demonstrate cross session continuity, an attendee asks in a new window, \textit{``I remember in our last taxi analysis, one Vendor had a tip percentage over 2500\%. Can you check if that specific trip was paid in cash or credit? I forgot to check.''} DataClaw utilizes \texttt{memory\_search} to retrieve prior findings from \texttt{MEMORY.md}, seamlessly extending the analytical thread. Furthermore, we demonstrate runtime extensibility by dropping a custom \texttt{experiment-tracker} Skill into the directory. DataClaw instantly registers the new capabilities without a restart, proving its adaptability to specific workflows.

\textbf{3. Operational Characteristics.} 
The live walkthrough highlights DataClaw's robust local performance. Typical tasks average 45 seconds to complete with several tool invocations. The system efficiently handles memory compaction at an 80\% context threshold and can sustain over 50 concurrent agent instances across various LLM backends. User can also configure the threshold on console to see how the memory compaction works in real time. This demonstrates that DataClaw is not only a powerful analytical assistant but also a practical solution for real-world data tasks.

\section{Conclusion}
\label{sec:conclusion}
In this paper, we introduce DataClaw, an autonomous data agent that helps users tackle daily data tasks through natural language interactions within their familiar IM environments. By integrating a transparent ReAct reasoning architecture, a multi-tiered memory system, and a pluggable skill mechanism, DataClaw acts as a capable personal data assistant. Through our live demonstration, attendees experience how DataClaw transforms a simple request into a complete analytical workflow. Furthermore, by keeping all data, artifacts, and computation on local infrastructure, DataClaw offers a concrete demonstration that rigorous data governance and agent autonomy are not in conflict. 

\bibliographystyle{ACM-Reference-Format}
\citestyle{acmnumeric}
\bibliography{sample}

\end{document}